\documentclass[]{aa}  

\usepackage{graphicx}
\usepackage{fixltx2e}
\usepackage{txfonts,booktabs}
\begin{document}
   \title{Binaries discovered by the MUCHFUSS project:\\
    SDSS J162256.66$+$473051.1 - An eclipsing subdwarf B binary with a brown dwarf companion
   }

   \titlerunning{SDSS J162256.66$+$473051.1 - An eclipsing subdwarf B binary with a brown dwarf companion}
   \authorrunning{Schaffenroth et al.}

   \author{V. Schaffenroth \inst{1,2} \and  S.Geier \inst{3,1} \and U. Heber \inst{1} \and T. Kupfer \inst{4} \and E. Ziegerer \inst{1} \and C. Heuser \inst{1} \and L. Classen \inst{1} \and
             O. Cordes \inst{5}
          }

   \institute{Dr.\,Remeis-Observatory \& ECAP, Astronomical Institute, Friedrich-Alexander University Erlangen-N\"urnberg, Sternwartstr.~7, 96049
 Bamberg, Germany\\
              \email{veronika.schaffenroth@sternwarte.uni-erlangen.de}
         \and
         Institute for Astro- and Particle Physics, University of Innsbruck, Technikerstr. 25/8, 6020 Innsbruck, Austria
         \and
         European Southern Observatory, Karl-Schwarzschild-Str. 2, 85748 Garching, Germany
         \and
         Department of Astrophysics/IMAPP, Radboud University Nijmegen, P.O. Box
         9010, 6500 GL Nijmegen, The Netherlands
         \and
         Argelander Institute for Astronomy, Auf dem H\"ugel 71, D-53121 Bonn,Germany
             }

   \date{Received 07 January 2014/ Accepted 19 February 2014}

\abstract{Hot subdwarf B stars (sdBs) are core helium-burning stars located on the extreme horizontal branch. About half of the known sdB stars are found in close binaries. Their short orbital periods of 1.2 h to a few days suggest that they are post common-envelope systems. Eclipsing hot subdwarf binaries are rare, but important to determine the fundamental stellar parameters. Low-mass companions are identified by the reflection effect. In most cases the companion is a main sequence star near the stellar mass limit.\\
 Here we report the discovery of an eclipsing hot subdwarf binary SDSS J162256.66+473051.1 (J1622) of very short orbital period (0.0697 d), found in the course of the MUCHFUSS project. The lightcurve shows grazing eclipses and a prominent reflection effect. An analysis of the light- and radial velocity (RV) curves indicated a mass ratio of $q=$ 0.1325, an RV semiamplitude $K=47.2\rm\,km\,s^{-1}$, and an inclination of $i=72.33^\circ$. We show that a companion mass of 0.064 $M_{\rm \odot}$, well below the hydrogen-burning limit, is the most plausible solution, which implies a mass close to the canonical mass (0.47 $M_{\rm \odot}$) of the sdB star. Therefore, the companion is a brown dwarf, which not only survived the engulfment by the red-giant envelope, but also triggered its ejection, and enabled the sdB star to form.\\
 %J1622 adds further evidence that brown dwarfs may indeed significantly affect fate of their host star.
 The rotation of J1622 is expected to be tidally locked to the orbit. However, J1622 rotates too slowly ($v_{\rm rot}=74.5\pm 7\rm\,km\,s^{-1}$) to be  synchronized, challenging tidal interaction models.}
   \keywords{stars: subdwarfs, binaries: eclipsing, binaries: spectroscopic, stars: brown dwarfs, stars: fundamental parameters, stars: individual: SDSS J162256.66$+$473051.1
               }

   \maketitle
%
%________________________________________________________________

\section{Introduction}
Hot subdwarfs (sdBs) are core helium-burning stars with very thin hydrogen envelopes, that are not able to sustain hydrogen shell burning \citep{heber:2009}. To form such objects, the progenitor has to lose almost all of its hydrogen envelope. The high percentage of 50\% of close binaries \citep{maxted:2001,napi} suggests that these sdBs are formed via binary evolution, which is responsible for the required large mass loss on the red giant branch (RGB). Such sdB binaries are formed via a common envelope phase or by stable Roche lobe overflow. The other half of sdB stars, however, appears to be single. Hence, binary evolution seems to be irrelevant at first glance. Nevertheless, is it possible to form a single star through merging of the components of a binary star. To form a helium core burning object like an sdB star, requires two helium white dwarfs to merge and ignite helium burning. The binary components are driven into a merger by gravitational wave radiation \citep{webbink,it}. The merged object is expected to rotate rapidly. However, rotation velocities of single sdB stars are very slow \citep[$\rm <=10\,km/s$,][]{geier:2012}, which is at variance with the merger scenario. Therefore, the merger scenario may explain a few cases of
exceptionally fast rotators \citep{geier:2011_3,geier:2013_rot}. Moreover, binary population synthesis predicts a wide mass distribution around 0.52 $M_{\rm \odot}$ for the sdB, resulting from the merger channel \citep{han:2002,han:2003}. The empirical mass distribution derived by asteroseismology \citep{fontaine,vangrootel:epjwc}, however, shows a sharp peak at $0.47 M_\odot$ for single sdB stars, which is inconsistent with the prediction of the merger scenario.

An alternative scenario proposes the engulfment and possible destruction of a substellar object within a common envelope (CE) as formation channel for single sdB stars \citep{soker,nt}. Close binary sdB stars with very low-mass stellar companions are known for decades \citep{kilkenny:1978,menzies:1986}. When eclipsing, such systems are named HW Vir stars. Most of the dozen known HW Vir systems have periods as short as 0.1 days. Hence, they must have undergone a common envelope and spiral-in phase. Binary population synthesis shows that sdBs in such post-common envelope systems should have masses around $0.47\,M_{\odot}$, the so-called canonical mass \citep{han:2002,han:2003}. The companions in HW Vir type binaries have masses close to, but usually exceeding the nuclear-burning limit of $\simeq0.08\,M_{\odot}$ and are, therefore, very late M-dwarf stars.  

Whether a substellar companion would also be able to contribute enough energy and angular momentum to unbind a common envelope, has been under debate. The discovery of a brown dwarf orbiting an sdB star in the HW Vir type binary SDSS J082053.53+000843.4 demonstrated that substellar companions are able to form an sdB \citep{geier}. The subsequent discoveries of two Earth-sized bodies orbiting a single pulsating sdB, which might be the remnants of one or two more massive companion evaporated in the CE-phase \citep{charpinet:nature,bear}, provides further evidence that substellar companions play an important role in the formation of close binary and single sdB stars alike. 
One of the open questions is now, how massive the companion has to be, to trigger the mass loss of the red giant to form an sdB star. 

Here we report on the discovery of a short period eclipsing hot subdwarf with a substellar companion in the course of the MUCHFUSS project. The project Massive Unseen Companions to Hot Faint Underluminous Stars from SDSS (MUCHFUSS) aims at finding sdBs with compact companions, such as massive white dwarfs  (M  > 1.0 $M_{\odot}$), neutron stars, or black holes. Details about target selection and the scope of the project can be found in \citet{geier:2011_2}. As one of two selection criteria, sdBs with radial velocity (RV) variations on timescales of half an hour or less were selected from the Sloan Digital Sky Survey (SDSS). These targets are either sdBs with compact, massive companions and comparatively long orbital periods, or short-period binaries with low-mass companions like the system described here.

To distinguish low-mass stellar or substellar companions from white dwarfs, and other more compact objects, we also started a photometric follow-up. As HW Vir systems have characteristic lightcurves, this turned out to be the ideal approach to find such binaries. A low-mass, cool companion can be identified due to the so-called reflection effect, which results from the large difference in temperature of both components. The companion is heated up on one side, and, therefore, a sinusoidal variation of the flux with the phase can be observed, as the heated hemisphere emits more flux. This effect is also visible at rather low inclinations, where no eclipses are observed. If eclipses are present, a combined spectroscopic and photometric analysis allows us to put tight constraints on the binary parameters. In the course of the MUCHFUSS project we already discovered three new HW Vir systems \citep{geier,tuc_schaff} and a pulsating sdB in a reflection effect binary \citep{ostenson:2013}.

In this paper we present the observations and analysis of a HW Vir star, discovered by the MUCHFUSS project. We describe the spectroscopic and photometric observations in Sect. 2 and their analysis in Sect. 3 (spectroscopy) and 4 (photometry). Evidence for the brown dwarf nature of the companion is given in Sect. 5 and the lack of synchronisation of the sdB star is discussed in Sect. 6. Finally, we conclude and present suggestions, how to improve the mass determination.

%__________________________________________________________________

\section{Observations}

\subsection{Spectroscopy}

The SDSS spectra of SDSS J162256.66+473051.1 (J1622 for short, also known as  PG1621+476, g'=15.96 mag), showed a radial velocity shift of 100 $\rm km\,s^{-1}$ within 1.45 h. Therefore, the star was selected as a high priority target for follow-up. 18 spectra were taken with the ISIS instrument at the William Herschel Telescope on La Palma, on 24 to 27 August 2009 at medium resolution (R$\sim$4000). Additional 64 spectra with the same resolution were obtained with the TWIN spectrograph at the 3.5 m telescope at the Calar Alto Observatory, Spain, on 25 to 28 May 2012. Moreover, 10 higher resolution spectra (R$\sim$8000) were observed with ESI at the Keck Telescope, Hawaii, on 13 July 2013. The TWIN and ISIS data were reduced with the MIDAS package distributed by the European Southern Observatory (ESO). The ESI data was reduced with the pipeline {\bf Makee\footnote{\url{http://www.astro.caltech.edu/~tb/ipac_staff/tab/makee/ }}.}

\subsection{Photometry}
\label{photo}
As the RV-curve from the ISIS spectra showed a short period of only 0.069 d, a photometric follow-up was started. J1622 was observed with the Bonn University Simultaneous Camera  (BUSCA) at the 2.2 m telescope at the Calar Alto Observatory. BUSCA \citep{reif} can observe in 4 bands simultaneously. We did not use any filters, but instead the intrinsic transmission curve given by the beam splitters, which divides the visible light into four bands $U_B$, $B_B$, $R_B$, and $I_B$ (and means no light loss). Four {sets of lightcurves}, each covering one orbit of J1622, were obtained on 12 June 2010, 29 September 2010, 28 February 2011, and 1 June 2011. They were reduced by using the aperture photometry package of IRAF. As the comparison stars have different spectral types, we observed a long-term trend in the lightcurve with changing air mass due to the different wavelength dependency of atmospheric extinction, which can not be corrected for.
\begin{figure*}[t]
\centering
\includegraphics[width=0.495\linewidth]{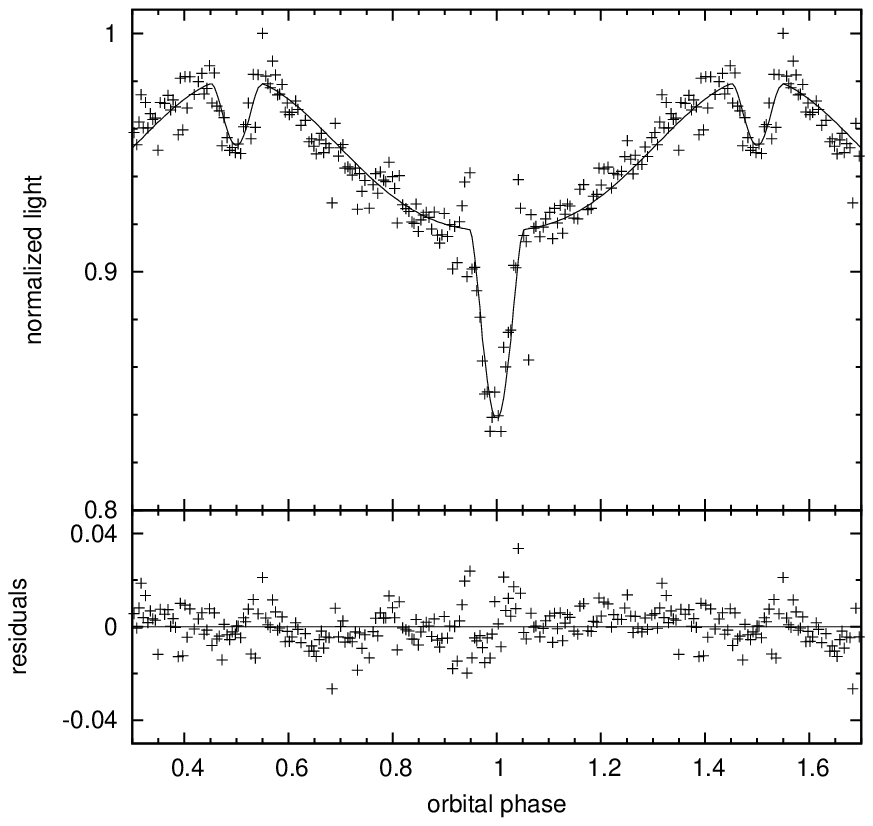}
\includegraphics[width=0.495\linewidth]{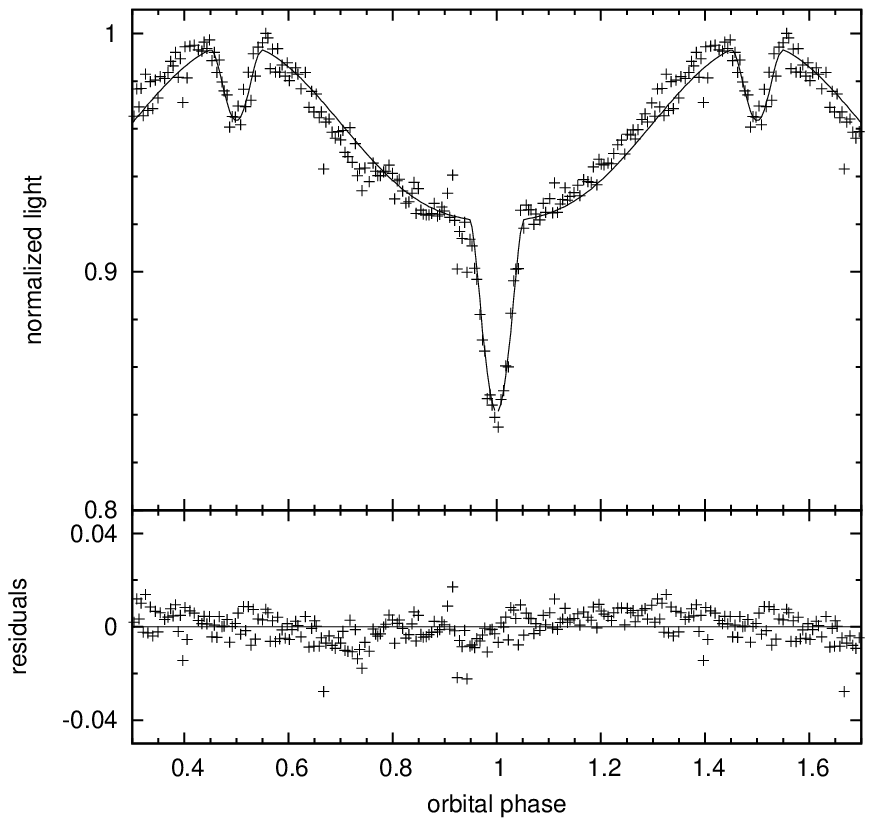}
\caption{Phased BUSCA lightcurve in $B_B$ and $R_B$ of J1622. The solid line demonstrates the best-fitting model. In the bottom panel the residuals can be seen.}
\label{lc}
\end{figure*}
%______________________________________________________________

\section{Spectroscopic analysis}

\subsection{Radial velocity curve}

\begin{figure}
\centering
\includegraphics[width=1.0\linewidth]{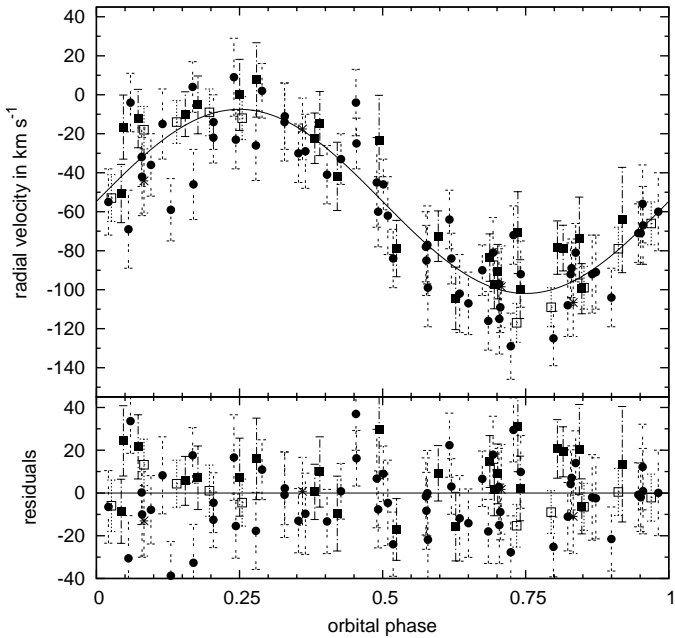}
\caption{Radial velocity plotted against orbital phase of J1622. The radial velocity was determined from the SDSS, ISIS, TWIN, and ESI spectra. All spectra were fitted together. The stars mark the SDSS spectra, the dots the TWIN spectra, the filled squares the ISIS spectra, and the open squares the ESI spectra. The errors are formal 1-$\sigma$ uncertainties. The lower panel shows the residuals.}
\label{rv}
\end{figure}

The radial velocities were measured by fitting a combination of Gaussians, Lorentzians, and polynomials to the Balmer and helium lines of all spectra and are given in Table~\ref{rv_measure}. Since the phase-shift between primary and secondary eclipses in the phased lightcurve (see Fig. \ref{lc}) is exactly 0.5, we know that the orbit of J1622 is circular. Therefore, sine curves were fitted to the RV data points in fine steps over a range of test periods. For each period, the $\chi^2$ of the best fitting sine curve was determined \citep[see][]{geier:2011}. All three datasets were fitted together. The orbit is well covered. Fig.  \ref{rv} shows the phased RV curve with the fit of the best solution. It gives a semiamplitude of $\rm K= 47.2 \pm 2.0 \,km\,s^{-1}$, a system velocity of $\rm \gamma= -54.7 \pm 1.5 \,km\,s^{-1}$, and a period of 0.0696859 $\pm$ 0.00003 d. The period is consistent with the period from the photometry (see Sect. \ref{period}), and the second shortest ever measured for an sdB binary. 

\subsection{Atmospheric parameters}

\begin{figure*}[t]
\centering
\includegraphics[angle=-90,width=0.495\linewidth]{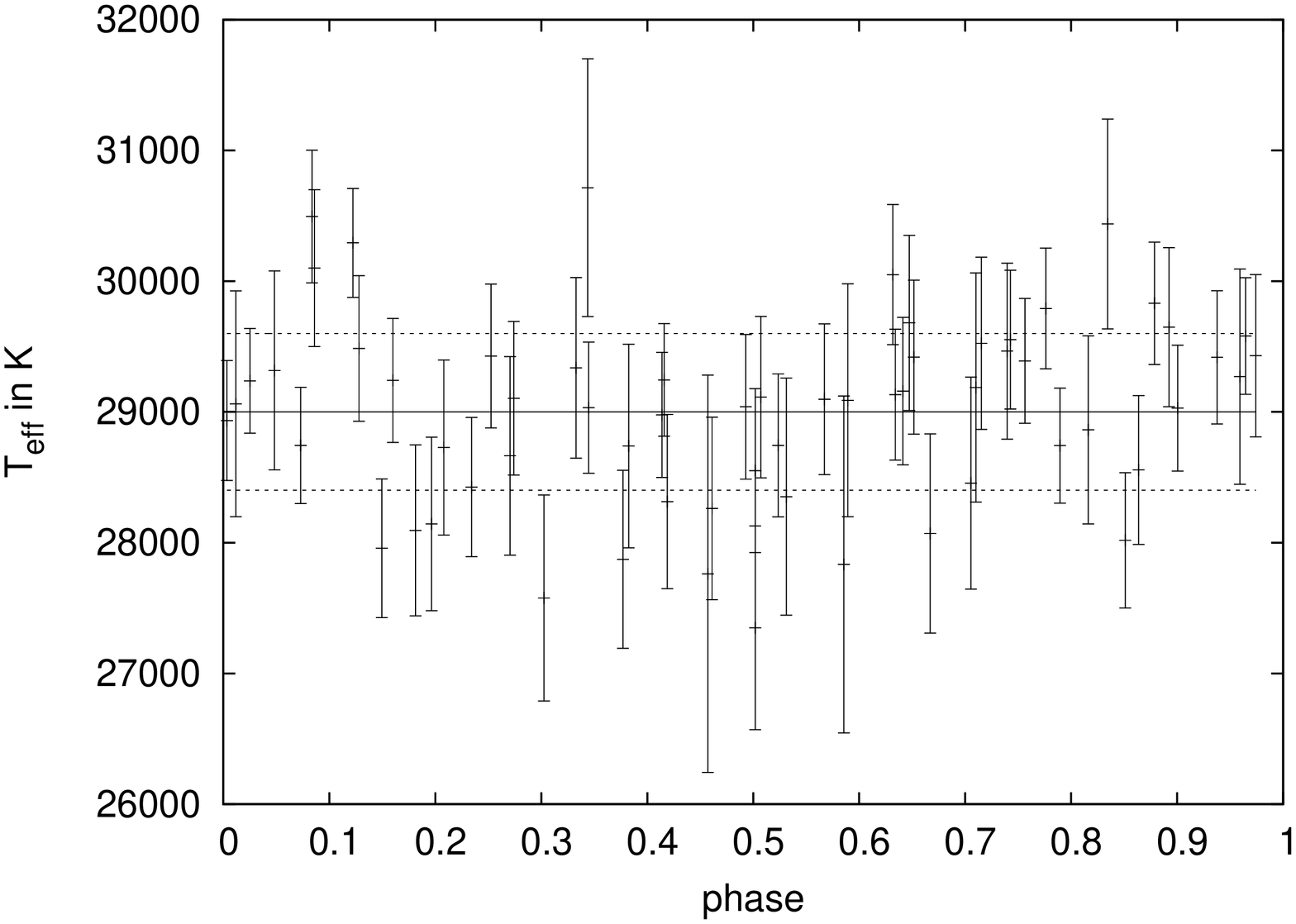}
\includegraphics[angle=-90,width=0.495\linewidth]{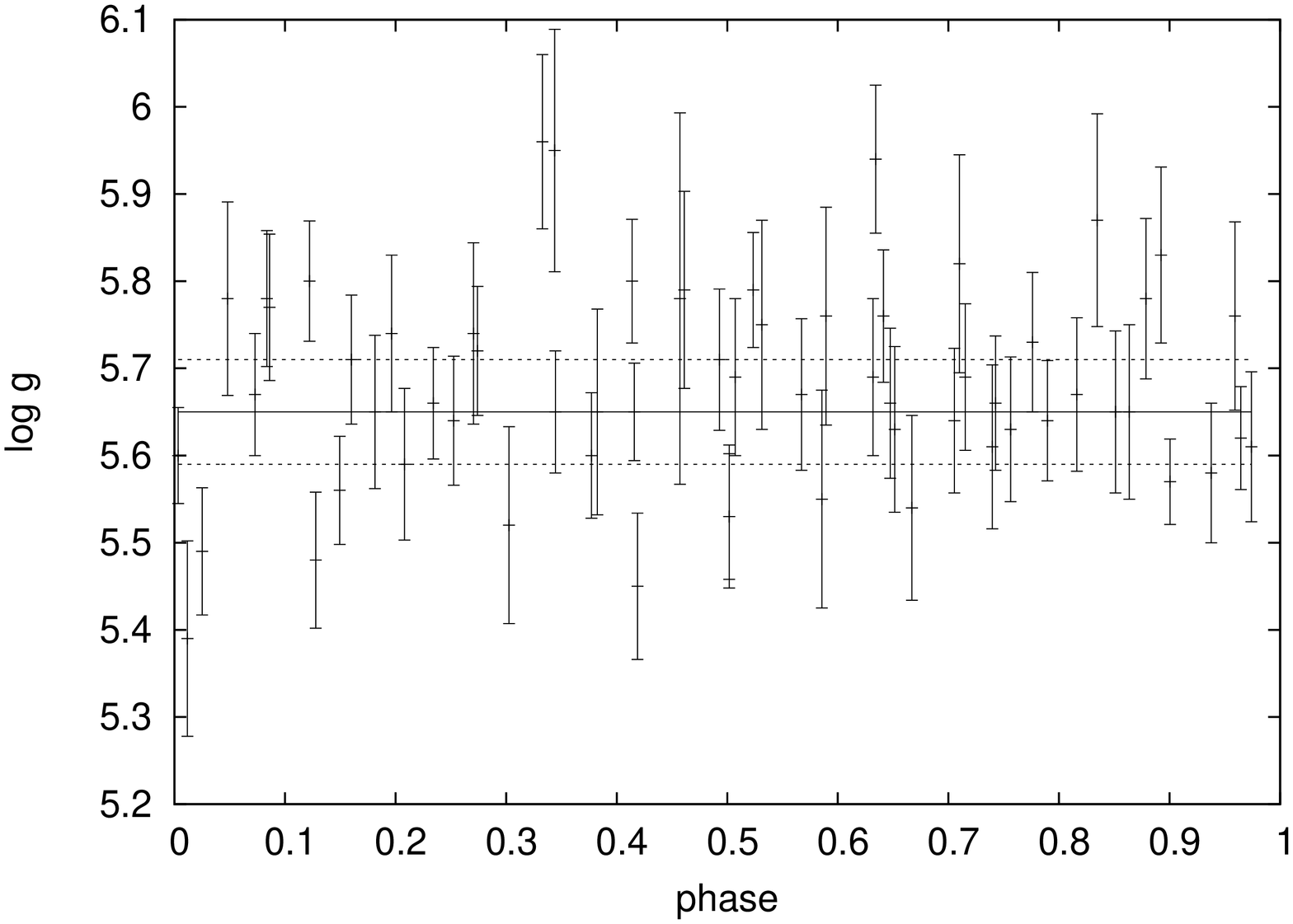}
\caption{Effective temperature and surface gravity plotted over the phase of J1622. $\rm T_{eff}$ and $\log{g}$ were determined from the TWIN spectra. The errors are statistical 1-$\sigma$ errors. The lines represent the parameters from the coadded spectrum together with the errors.}
\label{teff}
\end{figure*}

\begin{figure}

\centering
\includegraphics[angle=-90,width=1.0\linewidth]{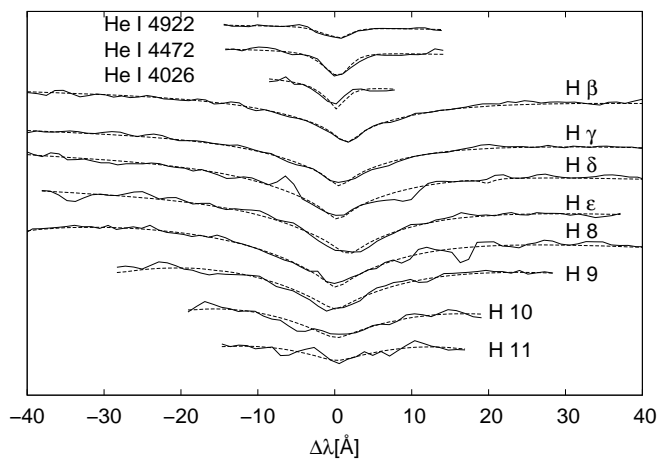}
\caption{Fit of the Balmer and helium lines in the coadded TWIN spectrum, the solid line shows the measurement and the dashed line shows the best fitting synthetic spectrum.}
\label{linien}
\end{figure}

The atmospheric parameters were determined by fitting synthetic spectra, that were calculated using LTE model atmospheres with solar metallicity and metal line blanketing \citep{heber:2000}, to the Balmer and helium lines using SPAS \citep{hirsch}. For some of the HW Vir stars and similar non-eclipsing systems, it was found that the atmospheric parameters seemed to vary with the phase \citep[e.g.][]{vs}, as the contribution of the companion to the spectrum varies with the phase. Therefore, all spectra were fitted separately. Fig. \ref{teff} shows the effective temperature and the surface gravity determined from the TWIN spectra plotted against orbital phase. No change with the orbital phase can be seen. Therefore, we coadded all 64 TWIN spectra and derived the atmospheric parameters ($\rm T_{eff}=29000\pm600\,{\rm K},\,\log{g}=5.65\pm0.06$, $\log{y}=-1.87 \pm 0.05$). In Fig. \ref{linien} the best fit to the Balmer and helium lines of the coadded spectrum is shown. 

\begin{figure}
\centering
\includegraphics[width=1.0\linewidth]{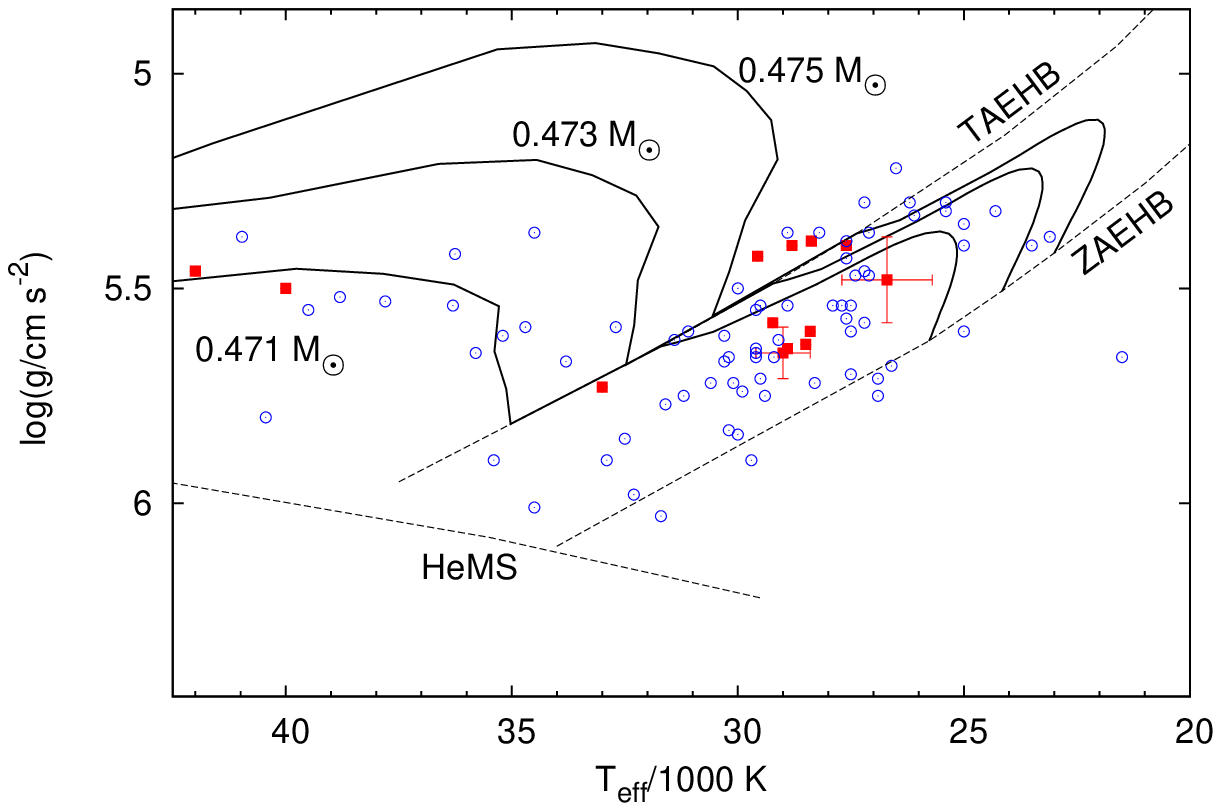}
\caption{$\rm T_{eff}-\log{g}$ diagram of the HW Vir systems. The solid lines are evolutionary tracks by \citet{Dorman:1993} for an sdB mass of 0.471, 0.473, and 0.475 $M_{\odot}$. The positions of J1622 and J0820 are indicated with crosses. The other squares mark the position of other HW Vir-like systems \citep{vangrootel, drechsel:2001, for:2010, geier, maxted:2002, klepp:2011, oestenson:2008, oestenson:2010, Wood:1999,almeida:2012,barlow:2012}. The open dots represent other sdB binaries from the literature.}
\label{hd}
\end{figure}

In Fig. \ref{hd} the position of J1622 in the $\rm T_{eff}-\log{g}$ diagram is compared to those of the known HW Vir systems and other sdB binaries. It is worthwhile to note that all of the HW Vir systems, but two, which have evolved off the EHB, have very similar atmospheric parameters. Those systems cluster in a distinct region of the $\rm T_{eff}-\log{g}$ diagram. Unlike the HW Vir stars other sdB binaries are distributed more or less uniformly across the extreme horizontal branch (see Fig. \ref{hd}). In this respect J1622 turns out to be a typical HW Vir system. 

\begin{figure}
\centering
\includegraphics[angle=-90,width=1.0\linewidth]{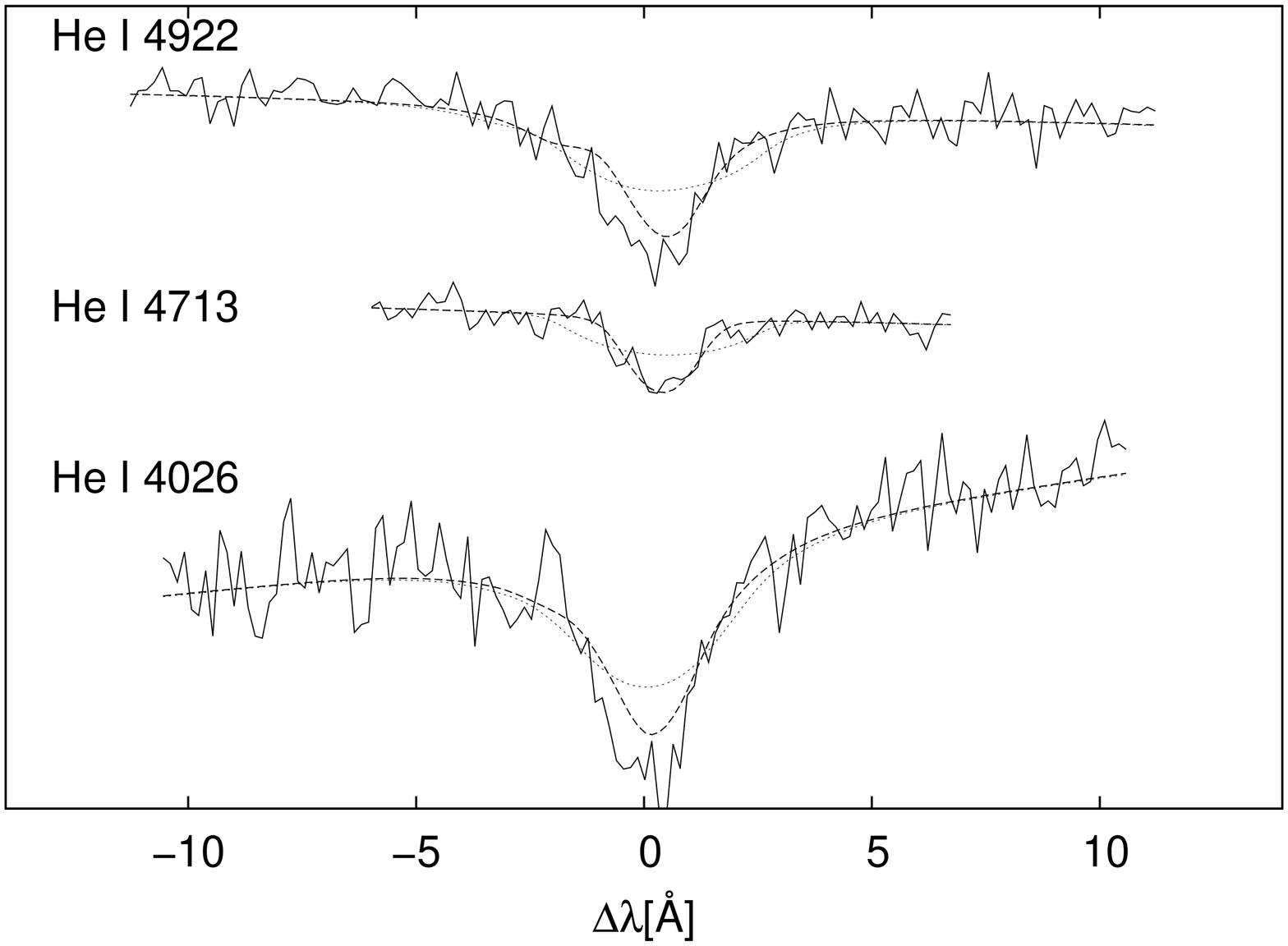}
\caption{Fit of the helium lines in the coadded ESI spectrum, the solid line shows the measurement and the dashed line shows the best fitting synthetic spectrum, the dotted line shows the line-broadening, if we assume synchronisation.}
\label{linien_esi}
\end{figure}

Due to their higher resolution, the ESI spectra are suitable to measure the rotational broadening of spectral lines of the sdB. We coadded all 10 spectra and determined the projected rotational velocity of the sdB primary by adding a rotational profile to the fit of the Balmer and helium lines. The other atmospheric parameters were kept fixed to the values determined from the TWIN spectra. The best fit for $\rm v_{\rm rot} \sin{i}$ = $\rm 71 \pm 7 {\rm \,km\,s^{-1}}$ is displayed in Fig. \ref{linien_esi}. 
Surprisingly, the projected rotational velocity is only about two thirds of the one expected for tidally locked rotation of the hot subdwarf primary. This issue will be further discussed in Sect.~\ref{rotation}.

%\begin{figure}
%\centering
%\includegraphics[angle=-90,width=1.0\linewidth]{logg}
%\caption{Surface gravity plotted over the phase of J1622. $\rm \log{g}$ was determined from the TWIN spectra. The errors are statistical one sigma errors. The lines represent the parameters from the coadded spectrum together with the errors.}
%\label{logg}
%\end{figure}
%______________________________________________________________

\section{Photometric analysis}

%\subsection{Ephemeris}
\label{period}
The BUSCA lightcurves clearly show a strong reflection effect and grazing eclipses, as can be seen in Fig. \ref{lc}. Unfortunately, the signal-to-noise of the $U_B$ and $I_B$  lightcurves is insufficient, so that only the $B_B$ and $R_B$ lightcurves are used for the analysis. 
The ephemeris was determined from the BUSCA lightcurves by fitting parabolas to the cores of the primary eclipses. The period was derived with the help of the Lomb-Scargle Algorithm \citep{press}.

The ephemeris of the primary minimum is given by\\
HJD = $2455359.58306(2) + 0.0697885(53) \cdot E$\\
 \citep[thereby is E the eclipse number, see e.g.][]{ drechsel:2001}

The phased lightcurves are  shown in Fig.~\ref{lc} . 
%Asymmetries with respect to the secondary eclipse become apparent, which might, however, not be real but caused by the observational shortcomings.  
%\subsection{Lightcurve analysis}
The lightcurve analysis was performed by using MORO \citep[MOdified ROche Program, see][]{drechsel:1995}, which calculates synthetic lightcurves, which were fitted to the observation. This lightcurve solution code is based on the Wilson-Devinney approach \citep{wilson:1971}, but uses a modified Roche model that considers the radiative pressure of hot binaries. More details of the analysis method are described in \citet{vs}. 

The main problem of the lightcurve analysis is the high number of parameters. To calculate the synthetic lightcurves, 12+5n (n is the number of lightcurves) not independent parameters are used. Therefore, strong degeneracies exist, in particular in the mass ratio, which is strongly correlated with the other parameters. The mass ratio is, therefore, fixed and solutions for different mass ratios are calculated. To resolve this degeneracy, it is, moreover, important to constrain as many parameters as possible from the spectroscopic analysis or theory. 

From the spectroscopic analysis we derive the effective temperature and the surface gravity of the sdB primary. Due to the early spectral type of the primary star the gravity darkening exponent can be fixed at $g_1=1$, as expected for radiative outer envelopes \citep{zeipel:1924}. For the cool convective companion, $g_2$ was set to 0.32 \citep{lucy:1967}. The linear limb darkening coefficients were extrapolated from the table of \citet{claret}.

To determine the quality of the lightcurve fit, the sum of the deviation of each point to the synthetic curve is calculated and the solution with the smallest sum is supposed to be the best solution. The difference between the solutions for the different mass ratios is unfortunately small, as expected. Therefore, we can not determine a unique solution from the lightcurve analysis alone and adopted the solution closest to the canonical mass for the sdB star. The corresponding results of the lightcurve analysis are given in Table \ref{asas} with errors determined with the bootstrapping method. The lightcurves in the $B_B$ and $R_B$ band are displayed in Fig.~\ref{lc} together with the best fit models for these parameters. 
The apparent asymmetries of the observed lightcurve can not be modelled, but we are not sure, if this effect is real or due to uncorrected long-term trends in the photometry  (see Sect. \ref{photo}). 
The parameters of the system resulting from the adopted solution together with the mass function are summarised in Table \ref{mass}. The errors result from error propagation of the errors of K, P and i.

%\begin{figure}
%\centering
%\includegraphics[width=1.0\linewidth]{lc_J082053}
%\caption{Phased BUSCA lightcurve in $R_B$ of J1622. The solid line demonstrates the best-fitting model. In the bottom panel the residuals can be seen.}
%\label{lc2}
%\end{figure}
%______________________________________________________________

\section{The brown dwarf nature of the companion}

\begin{figure}
\centering
\includegraphics[angle=-90,width=1.0\linewidth]{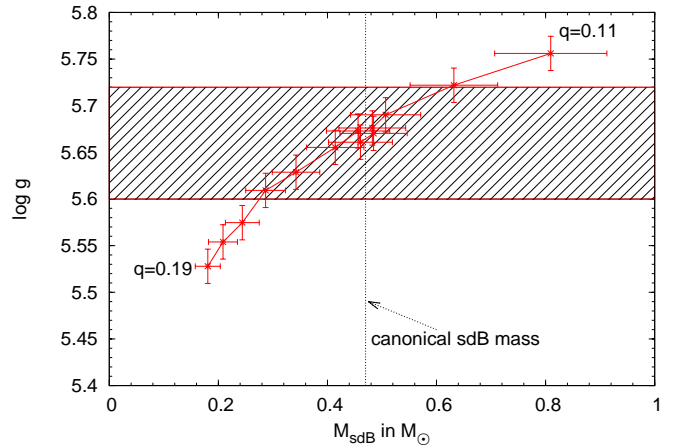}
\caption{Comparison of the photometric and spectroscopic surface gravity for the solutions with different mass ratio $q=0.11-0.19$ (marked by the error cross). The spectroscopic surface gravity with uncertainty is given by the shaded area.}
\label{sdB}
\end{figure}

From the semiamplitude of the radial velocity curve and the orbital period we can derive the masses and the radii of both components for each mass ratio. To constrain the solutions further, we first compared the photometric surface gravity, which can be derived from the mass and the radius, to the spectroscopic surface gravity. This is displayed in Fig. \ref{sdB}. The spectroscopic surface gravity is consistent with the lightcurve solution for sdB masses from 0.25 to 0.6 $M_{\rm \odot}$. It is, therefore, possible to find a self-consistent solution. This is not at all a matter of course, because in other cases such as AA Dor \citep{aador} gravity derived from photometry was found to be inconsistent with the spectroscopic result.

\begin{figure}
\centering
\includegraphics[width=1.0\linewidth]{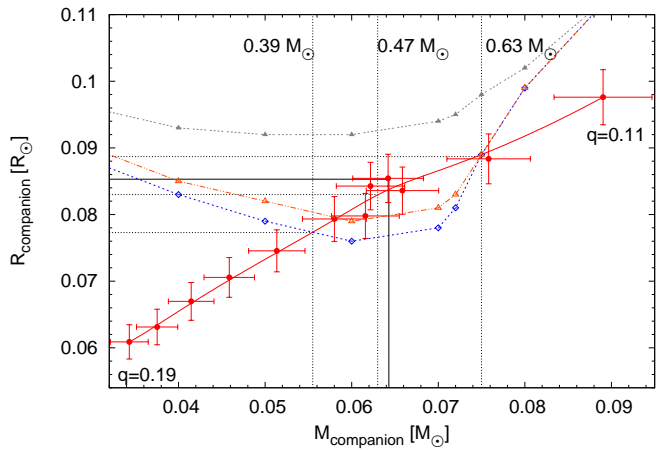}
\caption{Comparison of theoretical mass-radius relations of brown dwarfs by \citet{baraffe} for an age of 1 Gyr (filled triangles), 5 Gyrs (triangles) and 10 Gyrs (diamond) to results from the lightcure analysis.  Each error cross represents a solution from the lightcurve analysis for a different mass ratio ($q=0.11-0.19$). The dashed vertical lines mark different values of the corresponding sdB
masses. The solid lines mark the solution closest to the canonical mass for the sdB star of 0.47 $M_{\rm\odot}$, that was adopted.}
\label{bd}
\end{figure}

We also compared the radius of the companion with theoretical mass-radius relations for low-mass stars and brown dwarfs with ages of 1, 5, and 10 Gyrs, respectively \citep{baraffe}. As can be seen from Fig.~\ref{bd}, the measured mass-radius relation is well matched by theoretical predictions for stars 
$\gtrsim 3$ Gyrs for companion masses between 0.055 $M_{\rm \odot}$ and 0.075 $M_{\rm \odot}$. The corresponding mass range for the sdB star is from 0.39 $M_{\rm \odot}$ to 0.63 $M_{\rm \odot}$, calculated from the mass ratio.

 However, the companion is exposed to intense radiation of a quite luminous hot star, only 0.58 $R_{\odot}$ away, which could lead to an underestimate of the radius, if compared to
 non-irradiated models \citep{baraffe}. Such inflation effects have been found e.g. in the case of hot Jupiter exoplanets \citep[e.g.][]{udalski}, but also in the MS+BD binary CoRoT-15b \citep{bouchy}.  We can estimate the maximum inflation effect from theoretical mass-radius relations shown in Fig.~\ref{bd}. As can be seen from Fig.~\ref{bd}, inflation by more than 10\% can be excluded because otherwise none of the theoretical mass-radius relations would match the measured one, even if the star was as old as 10 Gyrs (the age of the Galactic disk).

If we assume an inflation of 5-10\%, the mass-radius relation for the companion would be in perfect agreement with the light curve solution for a companion with a mass of 0.064 $M_{\rm \odot}$ and a radius of 0.085 $R_{\rm \odot}$, and an age of $\sim$ 5-10 Gyrs. The corresponding mass of the sdB is close to the canonical sdB mass, which we therefore adopt for the sdB throughout the rest of the paper. As we calculated solutions for discrete q and, hence, discrete masses for the sdB and the companion, we adopted the solution closest to the canonical mass, which is also marked in Fig.~\ref{bd}.

\begin{table}
\caption{ Adopted lightcurve solution}
\label{asas}
\begin{tabular}{lcl}
\hline
\noalign{\smallskip}
\multicolumn{3}{l}{Fixed parameters:}\\
\noalign{\smallskip}
\hline
\noalign{\smallskip}
$q\,(=M_{2}/M_{1})$ & & $0.1325$\\
$T_{\rm eff}(1)$&[K]&\multicolumn{1}{l}{29000}\\
$g_1^b$&&\multicolumn{1}{l}{1.0}\\
$g_2^b$&&\multicolumn{1}{l}{0.32}\\
$x_1(B\_B)^c$&&\multicolumn{1}{l}{0.25}\\
$x_1(R\_B)^c$&&\multicolumn{1}{l}{0.20}\\
$\delta_2^d$&&\multicolumn{1}{l}{0.0}\\
\noalign{\smallskip}
\hline
\noalign{\smallskip}
\multicolumn{3}{l}{Adjusted parameters:}\\
\noalign{\smallskip}
\hline
\noalign{\smallskip}
$i$ & [$^{\rm \circ}$] & $72.33 \pm 1.11$\\
$T_{\rm eff}(2)$ & [K]& $2500 \pm 900$\\
$A_1^a$&&\multicolumn{1}{l}{$1.0\pm0.03$}\\
$A_2^a$ & & $0.9 \pm 0.2$\\
$\Omega_1^f$&&$3.646 \pm 0.17$\\
$\Omega_2^f$&&$2.359 \pm 0.054$\\
$\frac{L_1}{L_1+L_2}(B_B)^g$&&$0.99996 \pm 0.00077 $\\
$\frac{L_1}{L_1+L_2}(B_R)^g$&&$0.99984 \pm 0.00247 $\\
$\delta_1$&&$0.001 \pm 0.003$\\
$x_2(B_B)$&&$1.0 \pm 0.005$\\
$x_2(R_B)$&&$1.0 \pm 0.005$\\
$l_3(B_B)^f$&&\multicolumn{1}{l}{0.0}\\
$l_3(R_B)^f$&&\multicolumn{1}{l}{$0.045 \pm 0.008$}\\
\noalign{\smallskip}
\hline
\noalign{\smallskip}
\multicolumn{3}{l}{Roche radii$^h$:}\\
\noalign{\smallskip}
\hline
\noalign{\smallskip}
$r_1$(pole)&[a]&$0.284 \pm 0.013 $\\
$r_1$(point)&[a]&$0.290 \pm 0.015 $\\
$r_1$(side)&[a]&$0.290 \pm 0.014 $\\
$r_1$(back)&[a]&$0.290 \pm 0.014 $\\
\noalign{\smallskip}
$r_2$(pole)&[a]&$0.142 \pm 0.009$\\
$r_2$(point)&[a]&$0.150 \pm 0.011 $\\
$r_2$(side)&[a]&$0.144 \pm 0.009 $\\
$r_2$(back)&[a]&$0.149 \pm 0.011 $\\
\noalign{\smallskip}
\hline
\end{tabular}\\
\tablefoot{\\
$^{a}$ Bolometric albedo\\
$^{b}$ Gravitational darkening exponent\\
$^{c}$ Linear limb darkening coefficient; taken from \citet{claret} \\
$^{d}$ Radiation pressure parameter, see \citet{drechsel:1995}\\
$^{e}$ Fraction of third light at maximum\\
$^{f}$ Roche potentials\\
$^{g}$ Relative luminosity; $L_2$ is not independently adjusted, but recomputed from $r_2$ and $T_{\rm eff}$(2)\\
$^{h}$ Fractional Roche radii in units of separation of mass centres}
\end{table}

\begin{table}
\caption{Parameters of J1622}
\label{mass}
\centering
\begin{tabular}{c|c|c}
		\toprule
		\multicolumn{3}{c}{SDSS J162256.66$+$473051.1}\\ %$\equiv$ PG1621+4737}
		\toprule
		%%coordinates	&\multicolumn{2}{c}{16 22 56\, +47 30 51 (J2000.0)}\\
		%$\rm g'$&[mag]&16.0\\
		%\toprule
		i&$^\circ$&$72.33\pm1.11$\\
		 $M_{\rm sdB}$ & [$M_{\rm \sun}$] & $0.48\pm0.03$\\
		 $M_{\rm comp}$ & [$M_{\rm \sun}$] & $0.064\pm0.004$\\
		 $a$ & [$R_{\rm \sun}$] & $0.58\pm 0.02$\\
		$R_{\rm sdB}$ & [$R_{\rm \sun}$]& $0.168\pm 0.007$\\
		$R_{\rm comp}$ & [$R_{\rm \sun}$]&$0.085\pm 0.004$\\
		$\log{g}(\rm sdB,phot)$ & & $5.67\pm0.02$\\
	    $\log{g}(\rm sdB,spec)$ & & $5.65\pm0.06$\\
	    \bottomrule
\end{tabular}
\end{table}

%______________________________________________________________

\section{Synchronisation}
\label{rotation}

Most HW Vir systems have orbital separations as small as one solar radius. Hence, it is reasonable to expect that the rotation of both components is tidally locked to the orbit. Indeed the rotation rates of HW Vir and other objects of the class with similarly short periods (~0.1d) are found to be synchronised.
It is worthwhile to note  that the system PG\,1017$-$036, a reflection effect binary with almost the same parameters as J1622 ($P=0.072\,{\rm d}$, $K=51\,{\rm km\,s^{-1}}$, $T_{\rm eff}=30300\,{\rm K}$, $\log{g}=5.61$), has a measured $\rm v_{\rm rot} \sin{i}$ = $118\,{\rm km\,s^{-1}}$ \citep{maxted:2002}, fully consistent with synchronised rotation. However, J1622 ($P=0.0698\,{\rm d}$, $K=47\,{\rm km\,s^{-1}}$, $T_{\rm eff}=29000\,{\rm K}$, $\log{g}=5.65$) rotates with $\rm v_{\rm rot} \sin{i}$ = $\rm 71 {\rm \,km\,s^{-1}}$. With an inclination of $72.33^{\circ}$ this results in a rotational velocity of $\rm 74 {\rm \,km\,s^{-1}}$, which is only about two thirds of the rotational velocity expected for a synchronous rotation ($P_{\rm orbit}=P_{\rm rot}=\frac{2\pi R}{v_{\rm rot}}$).

The physical processes leading to synchronisation are not well understood in particular for stars with radiative envelopes such as sdB stars and rivalling theories \citep{zahn,tassoul} predict very different synchronisation timescales
\citep[for details see][]{geier:2010}. The synchronisation timescales increase strongly with orbital separation, hence with the orbital period of the system. 
In fact J1622 and PG\,1017$-$036 have amongst all known HW Vir systems the shortest periods and the highest probability for tidally locked rotation.
Therefore, it is surprising that J1622 apparently is not synchronised, while PG\,1017$-$036 is. Calculations even in the case of the less efficient mechanism \citep{zahn} predict that synchronisation should be established after $10^{5}\,{\rm yr}$, a time span much shorter than the EHB lifetime of $10^{8}\,{\rm yr}$.  

Evidence for non-synchronous rotation of sdB stars in reflection effect binaries was presented recently by
\citet{pablo:2011,pablo:2012}, who measured the rotational splittings of pulsation modes in three reflection effect sdB binaries observed by the Kepler space mission revealing that the subdwarf primaries rotate more slowly than synchronised. However, those binaries have rather long periods ($\sim 0.5\,{\rm d}$) and predicted synchronisation time scales are much longer than for J1622 and even exceed the EHB life time, if the least efficient synchronisation process \citep[see Fig. 19 in][]{geier:2010} is adopted. Hence, unlike for J1622, the non-synchronisation of those systems is not at variance with synchronisation theory.

It is quite unlikely that J1622 is too young for its rotation to be tidally locked to the orbit. Hence, we need to look for an alternative explanation. Tidal forces leading to circularisation and synchronisation, may lead to stable configurations,
in which the rotational and the orbital periods are in resonance, that is their ratio is that of integer numbers as observed for Mercury. Comparing the observed rotational period of J1622 to the orbital one, we find that the ratio is $0.607 \pm 0.065$, hence close to a 2 to 3 resonance. However, J1622 must have undergone a spiral-in phase through the common-envelope phase and it must be investigated, whether such a resonant configuration can persist through that dynamical phase.   

\section{Conclusions}

We performed a spectroscopic and photometric analysis of the eclipsing hot subdwarf binary J1622, that was found in the course of the MUCHFUSS project. The atmospheric parameters of the primary are typical for an sdB star in a HW Vir system. Mass-radius relations were derived for both components. The mass of the sdB star is constrained from spectroscopy (surface gravity) to 0.28 $M_{\odot}$ to 0.64\,$M_{\odot}$. The mass of the companion can be constrained by theoretical mass-radius relations to lie between 0.055\,$M_{\rm \odot}$ and 0.075\,$M_{\rm \odot}$, which implies that the sdB mass is between 0.39\,$M_{\rm \odot}$ and 0.63\,$M_{\rm \odot}$.

Assuming small corrections of about  5-10\% to the radius due to the inflation of the companion by the strong irradiation from the primary, a companion mass of  0.064\,$M_{\rm \odot}$ appears to be the most plausible choice resulting in a mass of the sdB close to the canonical mass of $0.47 M_{\rm\odot}$ star. Accordingly the companion is a substellar object. This is the second time a brown dwarf is found as a close companion to an sdB star. J1622 provides further evidence that substellar objects are able to eject a common envelope and form an sdB star. Finding more such systems, will help to constrain theoretical models \citep{soker,nt}.

An important result of the spectral analysis is that the sdB star is rotating slower than expected, if its rotation was locked to its orbit, as is observed for the very similar system 
PG\,1017$-$036. The non-synchronous rotation of J1622 is at variance with the predictions of tidal interaction models unless the sdB star is very young. The ratio of the rotational to the orbital period is close to a 2 to 3 resonance. However, it has to be investigated further, if such a configuration can survive the common envelope phase.

Future investigations should aim at the detection of spectral lines from the secondary, in order to measure its radial velocity, that is to turn the system into an  double-lined spectroscopic binary, which would allow to pin down the mass ratio. Emission lines of the companion's irradiated atmosphere have been detected in the sdOB system AA\,Dor \citep{aador} and most recently also for the prototype HW\,Vir \citep{tuc_maya}.  Since J1622 is quite compact and the irradiation of the companion strong, such emission lines might be detectable in high-resolution spectra of sufficient quality. Accurate photometry is needed to confirm or disprove any asymmetries in its lightcurve hinting at flows at the surface of the companion.

\begin{acknowledgements}
Based on observations collected at the Centro Astronómico Hispano Alemán (CAHA), operated jointly by the Max-Planck Institut für Astronomie and the Instituto de Astrofisica de Andalucia (CSIC) with BUSCA at 2.2m telescope, and with TWIN at the 3.5m telescope.\\
Based on observations made with the William Herschel Telescope operated on the island of La Palma by the Isaac Newton Group in the Spanish Observatorio del Roque de los Muchachos of the Instituto de Astrofísica de Canarias.\\
 Based on observations obtained at the W.M. Keck Observatory, which is operated as a scientific partnership among the California Institute of Technology, the University of California, and the National Aeronautics and Space Administration. The Observatory was made possible by the generous financial support of the W.M. Keck Foundation. The authors wish to recognise and acknowledge the very significant cultural role and reverence that the summit of Mauna Kea has always had within the indigenous Hawaiian community. We are most fortunate to have the opportunity to conduct observations from this mountain. \\
V.S. acknowledges funding by the Deutsches Zentrum f\"ur Luft- und Raumfahrt (grant 50 OR 1110) and by the Erika-Giehrl-Stiftung. T.K. acknowledges support from the Netherlands Research School of Astronomy (NOVA).
\end{acknowledgements}

\bibliography{aabib}

\begin{thebibliography}{54}
\expandafter\ifx\csname natexlab\endcsname\relax\def\natexlab#1{#1}\fi

\bibitem[{{Almeida} {et~al.}(2012){Almeida}, {Jablonski}, {Tello}, \&
  {Rodrigues}}]{almeida:2012}
{Almeida}, L.~A., {Jablonski}, F., {Tello}, J., \& {Rodrigues}, C.~V. 2012,
  \mnras, 423, 478

\bibitem[{{Baraffe} {et~al.}(2003){Baraffe}, {Chabrier}, {Barman}, {Allard}, \&
  {Hauschildt}}]{baraffe}
{Baraffe}, I., {Chabrier}, G., {Barman}, T.~S., {Allard}, F., \& {Hauschildt},
  P.~H. 2003, \aap, 402, 701

\bibitem[{{Barlow} {et~al.}(2013){Barlow}, {Kilkenny}, {Drechsel}, {Dunlap},
  {O'Donoghue}, {Geier}, {O'Steen}, {Clemens}, {LaCluyze}, {Reichart},
  {Haislip}, {Nysewander}, \& {Ivarsen}}]{barlow:2012}
{Barlow}, B.~N., {Kilkenny}, D., {Drechsel}, H., {et~al.} 2013, \mnras, 430, 22

\bibitem[{{Bear} \& {Soker}(2012)}]{bear}
{Bear}, E. \& {Soker}, N. 2012, \apjl, 749, L14

\bibitem[{{Bouchy} {et~al.}(2011){Bouchy}, {Deleuil}, {Guillot}, {Aigrain},
  {Carone}, {Cochran}, {Almenara}, {Alonso}, {Auvergne}, {Baglin}, {Barge},
  {Bonomo}, {Bord{\'e}}, {Csizmadia}, {de Bondt}, {Deeg}, {D{\'{\i}}az},
  {Dvorak}, {Endl}, {Erikson}, {Ferraz-Mello}, {Fridlund}, {Gandolfi},
  {Gazzano}, {Gibson}, {Gillon}, {Guenther}, {Hatzes}, {Havel}, {H{\'e}brard},
  {Jorda}, {L{\'e}ger}, {Lovis}, {Llebaria}, {Lammer}, {MacQueen}, {Mazeh},
  {Moutou}, {Ofir}, {Ollivier}, {Parviainen}, {P{\"a}tzold}, {Queloz}, {Rauer},
  {Rouan}, {Santerne}, {Schneider}, {Tingley}, \& {Wuchterl}}]{bouchy}
{Bouchy}, F., {Deleuil}, M., {Guillot}, T., {et~al.} 2011, \aap, 525, A68

\bibitem[{{Charpinet} {et~al.}(2011){Charpinet}, {Fontaine}, {Brassard},
  {Green}, {Van Grootel}, {Randall}, {Silvotti}, {Baran}, {{\O}stensen},
  {Kawaler}, \& {Telting}}]{charpinet:nature}
{Charpinet}, S., {Fontaine}, G., {Brassard}, P., {et~al.} 2011, \nat, 480, 496

\bibitem[{{Claret} \& {Bloemen}(2011)}]{claret}
{Claret}, A. \& {Bloemen}, S. 2011, \aap, 529, A75

\bibitem[{{Dorman} {et~al.}(1993){Dorman}, {Rood}, \&
  {O'Connell}}]{Dorman:1993}
{Dorman}, B., {Rood}, R.~T., \& {O'Connell}, R.~W. 1993, APJ, 419, 596

\bibitem[{{Drechsel} {et~al.}(1995){Drechsel}, {Haas}, {Lorenz}, \&
  {Gayler}}]{drechsel:1995}
{Drechsel}, H., {Haas}, S., {Lorenz}, R., \& {Gayler}, S. 1995, A{\rm \&}A,
  294, 723

\bibitem[{{Drechsel} {et~al.}(2001){Drechsel}, {Heber}, {Napiwotzki},
  {{\O}stensen}, {Solheim}, {Johannessen}, {Schuh}, {Deetjen}, \&
  {Zola}}]{drechsel:2001}
{Drechsel}, H., {Heber}, U., {Napiwotzki}, R., {et~al.} 2001, A{\rm \&}A, 379,
  893

\bibitem[{{Fontaine} {et~al.}(2012){Fontaine}, {Brassard}, {Charpinet},
  {Green}, {Randall}, \& {Van Grootel}}]{fontaine}
{Fontaine}, G., {Brassard}, P., {Charpinet}, S., {et~al.} 2012, \aap, 539, A12

\bibitem[{{For} {et~al.}(2010){For}, {Green}, {Fontaine}, {Drechsel}, {Shaw},
  {Dittmann}, {Fay}, {Francoeur}, {Laird}, {Moriyama}, {Morris},
  {Rodr{\'{\i}}guez-L{\'o}pez}, {Sierchio}, {Story}, {Strom}, {Wang}, {Adams},
  {Bolin}, {Eskew}, \& {Chayer}}]{for:2010}
{For}, B.-Q., {Green}, E.~M., {Fontaine}, G., {et~al.} 2010, APJ, 708, 253

\bibitem[{{Geier} {et~al.}(2011{\natexlab{a}}){Geier}, {Classen}, \&
  {Heber}}]{geier:2011_3}
{Geier}, S., {Classen}, L., \& {Heber}, U. 2011{\natexlab{a}}, \apjl, 733, L13

\bibitem[{{Geier} \& {Heber}(2012)}]{geier:2012}
{Geier}, S. \& {Heber}, U. 2012, \aap, 543, A149

\bibitem[{{Geier} {et~al.}(2013){Geier}, {Heber}, {Heuser}, {Classen},
  {O'Toole}, \& {Edelmann}}]{geier:2013_rot}
{Geier}, S., {Heber}, U., {Heuser}, C., {et~al.} 2013, \aap, 551, L4

\bibitem[{{Geier} {et~al.}(2010){Geier}, {Heber}, {Podsiadlowski}, {Edelmann},
  {Napiwotzki}, {Kupfer}, \& {M{\"u}ller}}]{geier:2010}
{Geier}, S., {Heber}, U., {Podsiadlowski}, P., {et~al.} 2010, \aap, 519, A25

\bibitem[{{Geier} {et~al.}(2011{\natexlab{b}}){Geier}, {Hirsch}, {Tillich},
  {Maxted}, {Bentley}, {{\O}stensen}, {Heber}, {G{\"a}nsicke}, {Marsh},
  {Napiwotzki}, {Barlow}, \& {O'Toole}}]{geier:2011_2}
{Geier}, S., {Hirsch}, H., {Tillich}, A., {et~al.} 2011{\natexlab{b}}, \aap,
  530, A28

\bibitem[{{Geier} {et~al.}(2011{\natexlab{c}}){Geier}, {Maxted}, {Napiwotzki},
  {{\O}stensen}, {Heber}, {Hirsch}, {Kupfer}, {M{\"u}ller}, {Tillich},
  {Barlow}, {Oreiro}, {Ottosen}, {Copperwheat}, {G{\"a}nsicke}, \&
  {Marsh}}]{geier:2011}
{Geier}, S., {Maxted}, P.~F.~L., {Napiwotzki}, R., {et~al.} 2011{\natexlab{c}},
  A\rm \&A, 526, A39+

\bibitem[{{Geier} {et~al.}(2011{\natexlab{d}}){Geier}, {Schaffenroth},
  {Drechsel}, {Heber}, {Kupfer}, {Tillich}, {{\O}stensen}, {Smolders},
  {Degroote}, {Maxted}, {Barlow}, {G{\"a}nsicke}, {Marsh}, \&
  {Napiwotzki}}]{geier}
{Geier}, S., {Schaffenroth}, V., {Drechsel}, H., {et~al.} 2011{\natexlab{d}},
  APJ, 731, L22+

\bibitem[{{Han} {et~al.}(2003){Han}, {Podsiadlowski}, {Maxted}, \&
  {Marsh}}]{han:2003}
{Han}, Z., {Podsiadlowski}, P., {Maxted}, P.~F.~L., \& {Marsh}, T.~R. 2003,
  MNRAS, 341, 669

\bibitem[{{Han} {et~al.}(2002){Han}, {Podsiadlowski}, {Maxted}, {Marsh}, \&
  {Ivanova}}]{han:2002}
{Han}, Z., {Podsiadlowski}, P., {Maxted}, P.~F.~L., {Marsh}, T.~R., \&
  {Ivanova}, N. 2002, MNRAS, 336, 449

\bibitem[{{Heber}(2009)}]{heber:2009}
{Heber}, U. 2009, ARA{\rm \&}A, 47, 211

\bibitem[{{Heber} {et~al.}(2000){Heber}, {Reid}, \& {Werner}}]{heber:2000}
{Heber}, U., {Reid}, I.~N., \& {Werner}, K. 2000, \aap, 363, 198

\bibitem[{Hirsch(2009)}]{hirsch}
Hirsch, H. 2009, Phd thesis, Friedrich Alexander Universit\"at Erlangen
  N\"urnberg

\bibitem[{{Iben} \& {Tutukov}(1984)}]{it}
{Iben}, Jr., I. \& {Tutukov}, A.~V. 1984, \apjs, 54, 335

\bibitem[{{Kilkenny} {et~al.}(1978){Kilkenny}, {Hilditch}, \&
  {Penfold}}]{kilkenny:1978}
{Kilkenny}, D., {Hilditch}, R.~W., \& {Penfold}, J.~E. 1978, \mnras, 183, 523

\bibitem[{{Klepp} \& {Rauch}(2011)}]{klepp:2011}
{Klepp}, S. \& {Rauch}, T. 2011, A\rm \&A, 531, L7+

\bibitem[{{Lucy}(1967)}]{lucy:1967}
{Lucy}, L.~B. 1967, Zeitschrift f\"ur Astrophysik, 65, 89

\bibitem[{{Maxted} {et~al.}(2001){Maxted}, {Heber}, {Marsh}, \&
  {North}}]{maxted:2001}
{Maxted}, P.~f.~L., {Heber}, U., {Marsh}, T.~R., \& {North}, R.~C. 2001, MNRAS,
  326, 1391

\bibitem[{{Maxted} {et~al.}(2002){Maxted}, {Marsh}, {Heber}, {Morales-Rueda},
  {North}, \& {Lawson}}]{maxted:2002}
{Maxted}, P.~F.~L., {Marsh}, T.~R., {Heber}, U., {et~al.} 2002, MNRAS, 333, 231

\bibitem[{{Menzies} \& {Marang}(1986)}]{menzies:1986}
{Menzies}, J.~W. \& {Marang}, F. 1986, in IAU Symposium, Vol. 118,
  Instrumentation and Research Programmes for Small Telescopes, ed. J.~B.
  {Hearnshaw} \& P.~L. {Cottrell}, 305

\bibitem[{{Napiwotzki} {et~al.}(2004){Napiwotzki}, {Karl}, {Lisker}, {Heber},
  {Christlieb}, {Reimers}, {Nelemans}, \& {Homeier}}]{napi}
{Napiwotzki}, R., {Karl}, C.~A., {Lisker}, T., {et~al.} 2004, \apss, 291, 321

\bibitem[{{Nelemans} \& {Tauris}(1998)}]{nt}
{Nelemans}, G. \& {Tauris}, T.~M. 1998, \aap, 335, L85

\bibitem[{{{\O}stensen} {et~al.}(2013){{\O}stensen}, {Geier}, {Schaffenroth},
  Telting, Bloemen, Nemeth, Beck, Papics, Tillich, Ziegerer, Machado,
  Littlefair, Dhillon, Aerts, Heber, Maxted, G¨ansicke, \&
  Marsh}]{ostenson:2013}
{{\O}stensen}, R.~H., {Geier}, S., {Schaffenroth}, V., {et~al.} 2013, \aap,
  accepted

\bibitem[{{{\O}stensen} {et~al.}(2010){{\O}stensen}, {Green}, {Bloemen},
  {Marsh}, {Laird}, {Morris}, {Moriyama}, {Oreiro}, {Reed}, {Kawaler}, {Aerts},
  {Vu{\v c}kovi{\'c}}, {Degroote}, {Telting}, {Kjeldsen}, {Gilliland},
  {Christensen-Dalsgaard}, {Borucki}, \& {Koch}}]{oestenson:2010}
{{\O}stensen}, R.~H., {Green}, E.~M., {Bloemen}, S., {et~al.} 2010, \mnras,
  408, L51

\bibitem[{{{\O}stensen} {et~al.}(2008){{\O}stensen}, {Oreiro}, {Hu},
  {Drechsel}, \& {Heber}}]{oestenson:2008}
{{\O}stensen}, R.~H., {Oreiro}, R., {Hu}, H., {Drechsel}, H., \& {Heber}, U.
  2008, in Astronomical Society of the Pacific Conference Series, Vol. 392, Hot
  Subdwarf Stars and Related Objects, ed. {U.~Heber, C.~S.~Jeffery, \&
  R.~Napiwotzki}, 221--+

\bibitem[{{Pablo} {et~al.}(2011){Pablo}, {Kawaler}, \& {Green}}]{pablo:2011}
{Pablo}, H., {Kawaler}, S.~D., \& {Green}, E.~M. 2011, \apjl, 740, L47

\bibitem[{{Pablo} {et~al.}(2012){Pablo}, {Kawaler}, {Reed}, {Bloemen},
  {Charpinet}, {Hu}, {Telting}, {{\O}stensen}, {Baran}, {Green}, {Hermes},
  {Barclay}, {O'Toole}, {Mullally}, {Kurtz}, {Christensen-Dalsgaard},
  {Caldwell}, {Christiansen}, \& {Kinemuchi}}]{pablo:2012}
{Pablo}, H., {Kawaler}, S.~D., {Reed}, M.~D., {et~al.} 2012, \mnras, 422, 1343

\bibitem[{{Press} \& {Rybicki}(1989)}]{press}
{Press}, W.~H. \& {Rybicki}, G.~B. 1989, \apj, 338, 277

\bibitem[{{Reif} {et~al.}(1999){Reif}, {Bagschik}, {de Boer}, {Schmoll},
  {Mueller}, {Poschmann}, {Klink}, {Kohley}, {Heber}, \& {Mebold}}]{reif}
{Reif}, K., {Bagschik}, K., {de Boer}, K.~S., {et~al.} 1999, in Society of
  Photo-Optical Instrumentation Engineers (SPIE) Conference Series, Vol. 3649,
  Society of Photo-Optical Instrumentation Engineers (SPIE) Conference Series,
  ed. M.~M. {Blouke} \& G.~M. {Williams}, 109--120

\bibitem[{{Schaffenroth} {et~al.}(2013{\natexlab{a}}){Schaffenroth}, {Geier},
  {Barbu-Barna}, {Heber}, Kupfer, \& Cordes}]{tuc_schaff}
{Schaffenroth}, V., {Geier}, S., {Barbu-Barna}, I., {et~al.}
  2013{\natexlab{a}}, Astronomical Society of the Pacific Conference Series, in
  press

\bibitem[{{Schaffenroth} {et~al.}(2013{\natexlab{b}}){Schaffenroth}, {Geier},
  {Drechsel}, {Heber}, {Wils}, {{\O}stensen}, {Maxted}, \& {di Scala}}]{vs}
{Schaffenroth}, V., {Geier}, S., {Drechsel}, H., {et~al.} 2013{\natexlab{b}},
  \aap, 553, A18

\bibitem[{{Soker}(1998)}]{soker}
{Soker}, N. 1998, \aj, 116, 1308

\bibitem[{{Tassoul} \& {Tassoul}(1992)}]{tassoul}
{Tassoul}, M. \& {Tassoul}, J.-L. 1992, \apj, 395, 604

\bibitem[{{Udalski} {et~al.}(2008){Udalski}, {Pont}, {Naef}, {Melo}, {Bouchy},
  {Santos}, {Moutou}, {D{\'{\i}}az}, {Gieren}, {Gillon}, {Hoyer}, {Mayor},
  {Mazeh}, {Minniti}, {Pietrzy{\'n}ski}, {Queloz}, {Ramirez}, {Ruiz},
  {Shporer}, {Tamuz}, {Udry}, {Zoccali}, {Kubiak}, {Szyma{\'n}ski},
  {Soszy{\'n}ski}, {Szewczyk}, {Ulaczyk}, \& {Wyrzykowski}}]{udalski}
{Udalski}, A., {Pont}, F., {Naef}, D., {et~al.} 2008, \aap, 482, 299

\bibitem[{{Van Grootel} {et~al.}(2013{\natexlab{a}}){Van Grootel}, {Charpinet},
  {Brassard}, {Fontaine}, \& {Green}}]{vangrootel}
{Van Grootel}, V., {Charpinet}, S., {Brassard}, P., {Fontaine}, G., \& {Green},
  E.~M. 2013{\natexlab{a}}, \aap, 553, A97

\bibitem[{{Van Grootel} {et~al.}(2013{\natexlab{b}}){Van Grootel}, {Fontaine},
  {Charpinet}, {Brassard}, \& {Green}}]{vangrootel:epjwc}
{Van Grootel}, V., {Fontaine}, G., {Charpinet}, S., {Brassard}, P., \& {Green},
  E.~M. 2013{\natexlab{b}}, in European Physical Journal Web of Conferences,
  Vol.~43, European Physical Journal Web of Conferences, 4007

\bibitem[{{von Zeipel}(1924)}]{zeipel:1924}
{von Zeipel}, H. 1924, MNRAS, 84, 665

\bibitem[{{Vu{\v c}kovi{\'c}} {et~al.}(2008){Vu{\v c}kovi{\'c}}, {{\O}stensen},
  {Bloemen}, {Decoster}, \& {Aerts}}]{aador}
{Vu{\v c}kovi{\'c}}, M., {{\O}stensen}, R., {Bloemen}, S., {Decoster}, I., \&
  {Aerts}, C. 2008, in Astronomical Society of the Pacific Conference Series,
  Vol. 392, Hot Subdwarf Stars and Related Objects, ed. U.~{Heber}, C.~S.
  {Jeffery}, \& R.~{Napiwotzki}, 199

\bibitem[{{Vu{\v c}kovi{\'c}} {et~al.}(2013)}]{tuc_maya}
{Vu{\v c}kovi{\'c}}, M. {et~al.} 2013, Astronomical Society of the Pacific
  Conference Series, in press

\bibitem[{{Webbink}(1984)}]{webbink}
{Webbink}, R.~F. 1984, \apj, 277, 355

\bibitem[{{Wilson} \& {Devinney}(1971)}]{wilson:1971}
{Wilson}, R.~E. \& {Devinney}, E.~J. 1971, APJ, 166, 605

\bibitem[{{Wood} \& {Saffer}(1999)}]{Wood:1999}
{Wood}, J.~H. \& {Saffer}, R. 1999, MNRAS, 305, 820

\bibitem[{{Zahn}(1977)}]{zahn}
{Zahn}, J.-P. 1977, \aap, 57, 383

\end{thebibliography}
\bibliographystyle{aa}
\Online
\begin{appendix}
\setcounter{table}{0}
\renewcommand{\thetable}{A\arabic{table}}

%\section*{Radial velocities with errors of J1622}
\begin{table*}
\caption{Radial velocities with errors of J1622}
\label{rv_measure}
\centering
\begin{tabular}{lr}
\begin{tabular}{c|cc}
	\toprule
		\multicolumn{3}{c}{SDSS}\\\hline
		HJD&\multicolumn{2}{c}{RV in $\rm km\,s^{-1}$}\\\toprule
2452378.381481 &	-107   &     17\\                  
2452378.398900 &	-44    &    16 \\
2452378.418200 &	-18    &    16 \\
2452378.442390 &	-99    &    21 \\
	\toprule
		\multicolumn{3}{c}{WHT}\\\toprule
2455039.90179  &	-64    &    27 \\
2455068.90762 & 	-10     &   11  \\
2455068.92338 & 	-22     &   13 \\
2455068.94468 & 	-84     &   12 \\
2455069.85954  &	-79     &   12 \\
2455069.92083  &	-98     &   12 \\
2455069.93150  &	-100     &   13 \\
2455070.85307  &	-12     &   1 \\
2455070.86036 & 	-5     &   15  \\
2455070.86749 &		8     &	   19\\
2455070.87733  &	-42    &    18 \\
2455070.88454  &	-79    &    14 \\
2455070.89174  &	-104   &     16\\
2455071.93491  &	-73    &    13 \\
2455071.94215  &	-91    &    14 \\
2455071.94942  &	-78   &     14 \\
2455071.96599  &	-51    &    15 \\
2455071.96626  &	-17    &    16 \\
2455071.98042  &	0     &   18  \\
2455071.99018  &	-15    &    16 \\
2455071.99739  &	-23    &    23 \\
2455072.01429  &	-71    &    21 \\
2455072.01457  &	-100    &    15  \\
2455072.02178  &	-74    &    21  \\
	\toprule
		\multicolumn{3}{c}{TWIN}\\\toprule
2456073.37085  &	26   &		20  \\
2456073.33931  &	-4	&		13 \\
2456073.38513  & 	-62  &		20  \\   
2456073.39285  & 	-84  &		13  \\    
2456073.40042  & 	-72  &		15   \\   
2456073.40799  & 	-81  &		15   \\   
2456073.41591  & 	-71  &		16   \\   
2456073.42347 & 	 -4  &		15   \\   
2456073.43104 & 	 4    &		13   \\   
2456073.43950 & 	 2    &		14   \\   
2456073.44741 & 	 -41  &		15  \\    
2456073.45546 &		  -84  &	15  \\    
2456073.45951  &	 -85  &		18   \\   
2456073.46356  &	 -102  &	20  \\    
2456073.46762  &	 -81 	 &	18 \\     
2456074.37223  &	 -90  &		13 \\     
2456074.37296  &	 -116 &		15  \\    
2456074.38264  &	 -108	 &	16  \\    
2456074.38795  &	 -104 &		15  \\    
2456074.39366  &	 -60  &		20  \\       
	\hline
\end{tabular}\hspace{4cm}
\begin{tabular}{c|cc}
	\toprule
2456074.39883 &	 -69  &		20  \\    
2456074.40400  &	 -59 &		16  \\ 	
2456074.40918  &	 -14  &		14   \\   
2456074.41435  &	 -26  &		18  \\    
2456074.41952  &	 -30  &		15  \\    
2456074.42470  &	 -33  &		13  \\    
2456074.42987  &	 -46  &		13  \\    
2456074.43504  &	 -78  &		18   \\   
2456074.44022  &	-107 &		16  \\    
2456074.44539  &	 -129 &		17  \\    
2456074.45057 &		  -125 &	14 \\     
2456074.45574  &	 -91  &		19 \\     
2456074.46091  &	 -71  &		15  \\    
2456074.46607  &	 -55  &		17  \\    
2456074.47125  &	 -36  &		16  \\    
2456074.47643  &	 -46  &		18 \\     
2456074.48159  &	 -23  &		15 \\     
2456075.47452  &	 -60  &		18  \\    
2456075.48320   &	-64  &		15  \\    
2456075.49189  &	 -92  &		17 \\     
2456075.50058  &	 -92  &		20  \\    
2456075.51796  &	 -15  &		18 \\     
2456075.52665  &	 9   &		20  \\    
2456075.53535  &	 -29  &		19  \\    
2456075.54404  &	 -45  &		23 \\     
2456076.53468  &	 -109 &		13  \\    
2456076.54337  &	 -89  &		15  \\    
2456076.55206  &	 -67  &		20  \\    
2456076.56075  &	 -42  &		20  \\    
2456076.56944  &	 -22  &		13  \\    
2456076.57813  &	 -11   &	17  \\    
2456077.35340  &	 -25  &		13  \\    
2456077.36210  &	 -99  &		20  \\    
2456077.37079  &	 -115 &		18  \\    
2456077.37948  &	 -92  &		16  \\    
2456077.52761  &	 -56  &		20  \\    
2456077.53631  &	 -32  &		15   \\     
2456077.55370  &	 -14   &	      \\ 
2456077.56240  &	 -4   &		17   \\
2456077.57110  &	 -77  &		20    \\   
2456077.57980  &	 -97  &		18   \\    
	\toprule
		\multicolumn{3}{c}{ESI}\\\toprule
2456121.76284	 &	-117 &		10   \\
2456121.76704 &	-109 &		10   \\   
2456121.77103 &	-99 &		11   \\   
2456121.77522 &	-79 &		11   \\   
2456121.77922 &	-66 &		11   \\   
2456121.78318 &	-53 &		12   \\   
2456121.78715 &	-18 &		12   \\   
2456121.79114 &	-14 &		11   \\   
2456121.79511 &	-9 &		12   \\   
2456121.79910 &	-12 &		11   \\
	\hline
	\end{tabular}
\end{tabular}
\end{table*}
\end{appendix}

\end{document}